\documentstyle{aipproc}
 
\begin{document}
 
\title{Fermi Surfaces, Fermi Patches, and Fermi Arcs in High $T_c$
Superconductors}
\author{M. R. Norman}
\address{Materials Science Division, Argonne National Laboratory,
         Argonne, IL 60439}
\maketitle

\begin{abstract}
A defining property of metals is the existence of a Fermi surface: for 
two dimensions, a continuous contour in momentum space which separates 
occupied from unoccupied states.
In this paper, I discuss angle resolved photoemission data on the cuprate 
superconductor BSCCO and argue that it is not best thought of in this 
conventional picture.  Rather, the data are consistent with ``patches'' 
of finite area connected by more conventional ``arcs''.  Novel physics 
is associated with the patches, in that the states contained in a patch are 
dispersionless and thus interaction dominated.
In the pseudogap phase, the patches are gapped out, leaving the 
Fermi arcs disconnected.  This unusual situation may be the key to 
understanding the microscopic physics of the high temperature superconductors,
in that the pairing correlations are strongest in the patches, yet the 
superfluid density lives only on the arcs.
\end{abstract}

Perhaps no problem in recent history has captured the attention of the 
condensed matter physics community as much as that of the high 
temperature cuprate superconductors.  After over a decade of work, it has 
become clear that conventional metal physics alone is not capable in 
describing all of their properties.  In fact, it is strongly felt in a 
section of the community that new physics needs to be developed to fully 
solve the problem\cite{AND}.  One difficulty has been to properly define 
what the problem to be solved exactly is.  For instance, it is now generally
accepted by the community that conventional metal physics should not 
break down for a two dimensional metal which exhibits a free electron 
dispersion\cite{MOHIT}.  This can be contrasted with the unusual 
excitations associated with the fractional quantum hall effect, 
including that of the still debated half integral case, which arise due 
to the degeneracy of states in an applied magnetic field (Landau 
levels).  This points to the reasonable hypothesis that the novel
physics associated with the high temperature cuprate superconductors has 
something to do with a dramatic departure of their dispersion from that 
of free electron metals.

The experimental tool to employ in this regard is angle resolved 
photoemission spectroscopy (ARPES).  For two dimensional materials, if 
the impulse approximation is valid, then the ARPES intensity should be 
proportional to the product of the single particle spectral function and 
the Fermi function\cite{NK}.  The 2D condition is almost certainly satisfied 
in the cuprates, in that there is no evidence, particularly in BSCCO, for 
any c-axis dispersion in the data.  For the photon energies typically 
employed, there is also evidence that the impulse approximation is 
satisfied.  This is a result of the large number of available final 
states (due to the complexity of the crystal structure) coupled with the
short escape depth of the photoelectrons (which limits the time for the 
photoelectron to interact with the photohole).  As a 
consequence, the probe is surface sensitive, which has limited much of 
the most useful data to BSCCO, given its strongly two 
dimensional nature, with resulting cleaved surfaces characteristic of the 
bulk.

The importance of the single particle spectral function is self evident.  
It is the most fundamental quantity predicted by a many-body theory.  
Being proportional to the imaginary part of the Greens function, it can 
in principle be Kramers-Kronig transformed to yield the entire Greens 
function, which in turn defines the electron self-energy, the function 
that encapsulates the many-body physics of the material.  The self-energy 
implicitly contains the energy dispersion, which as argued above, has 
strong relevance to the nature of the ground and excited states of the system.

Although ARPES only reveals the occupied part of the spectral function, 
we are fortunate in the cuprates in that scanning tunneling microscope 
(STM) measurements in BSCCO yield spectra which strongly resemble that from
ARPES for the appropriate bias sign despite the fact that STM is a 
momentum averaged probe.  This indicates that the
tunneling matrix elements weight the spectra to certain regions of 
the Brillouin zone (to be identified later as the patches).  Thus, not 
only do these data provide a useful complement to ARPES data, with the 
additional advantage of spatial resolution and much higher energy 
resolution, they can be additionally exploited to obtain crucial information
concerning the unoccupied part of the spectral function.

With this introduction, we now turn to some general observations 
concerning ARPES data in Bi2212.  The identification of what the actual 
Fermi surface is in this material is still of some controversy.  This is 
easy to understand from early measurements of the Stanford 
group\cite{DESS93}.  In a very illuminating figure of a paper by that 
group (Fig.~2), they show those regions of the Brillouin zone where 
spectral weight is seen within 50 meV of the Fermi energy, which was 
comparable to their energy resolution.  They then interpreted this in 
terms of two conventional Fermi surfaces, an electron-like
surface centered at the $(0,0)$ point of the zone, and a hole-like surface
centered at the $(\pi,\pi)$ point of the zone.  This was conjectured to 
be due to the predicted bilayer splitting of the electronic structure due
to the two CuO planes per bilayer unit.  Subsequent measurements 
by the UIC-Argonne group\cite{DING96} were able to attribute the Fermi 
crossing of the presumed electron-like sheet along $(0,0)-(\pi,0)$ to
actually be
that of a ghost image of the Fermi surface due to diffraction of the 
outgoing photoelectrons by the superstructure associated with the BiO 
surface layer.  This led to what is now the mostly accepted picture of just a 
single hole-like surface centered around $(\pi,\pi)$, with the surprising 
result that somehow, the c-axis kinetic energy associated with bilayer 
splitting is missing in the dispersion of the spectral function.

I will now argue that this more or less conventional picture has been accepted
a bit too readily.  As stated above, what the authors of 
Ref.~\onlinecite{DESS93} actually show is a plot of the near Fermi energy 
spectral weight in the zone.  This plot is highly instructive.  What is 
seen is a mass of states in the vicinity of the $(\pi,0)$ points of the 
zone connected by more typical Fermi surface segments.  That is, the most 
natural interpretation of the data is not that one has a Fermi surface 
that represents a continuous contour in momentum space.  Rather, it 
appears that one has ``Fermi patches'' which occupy a finite {\it area} 
in momentum space, connected by ``Fermi arcs'' which resemble more 
typical behavior.  This observation is reinforced by directly looking at 
the raw spectra.  There, one sees that states in the ``patch'' region in 
the normal state are characterized by very broad linewidths which exhibit 
little dispersion.  In fact, the often quoted ``Fermi crossing'' along 
the $(\pi,0)-(\pi,\pi)$ direction is based not on dispersion, but rather 
on the drop in intensity of the spectra along this direction as would be 
expected from such a crossing.

This behavior becomes even more unusual in the superconducting
state.  Along the traditionally accepted hole-like Fermi 
contour, one sees a dispersion consistent with a superconducting order 
parameter of the $d_{x^2-y^2}$ form\cite{SHEN93,RAPID96}.  But in the ``patch'' 
region, one finds a narrow quasiparticle peak with little if any 
observable dispersion at an energy location given by the maximum of the 
d-wave gap\cite{NORM97}.  This is separated by a spectral dip from a 
higher binding energy feature, the ``hump'', which does in fact have 
strong dispersion in the ``patch''.  This dispersion becomes quite 
obvious in underdoped materials\cite{NEW99} where it begins to take on 
qualities reminiscent of the undoped magnetic insulator\cite{RONNING}.

It is of interest to note that the quasiparticle peak only forms below
$T_c$; that is, the Fermi liquid superconducting state appears to arise
from a non Fermi liquid normal state.  This is most dramatic in the
underdoped case, where a spectral gap already exists in the normal
state in the patches\cite{MARSHALL,DING,LOESER}.  In this case, one finds
a strongly incoherent spectrum above $T_c$, which loses any resemblance to
even a very broad peak as the doping is reduced.  Yet, despite this, one
finds the rather surprising presence of a sharp leading edge gap.  This is
a non-trivial finding which has yet to be reproduced by any theory which
purports
to explain the pseudogap.  Below $T_c$, a quasiparticle peak forms at this
gap edge.  This is somewhat reminiscent of the formation of a bound state
inside of a gap, which has suggested a composite nature for the incoherent
weight beyond the gap edge\cite{LAUGHLIN}.  Above $T_c$, the
spectral gap ``fills in'', leading to a more or less flat spectrum at
a characteristic temperature $T^*$\cite{NAT98}.  This filling in effect is
also infered from specific heat data\cite{LORAM}, and
directly observed as well by STM\cite{FISCHER} and c-axis optical
conductivity\cite{TIMUSK}.

The filling in seen in the patch region can be modeled by a very simple
self-energy of the form
$\Sigma=-i\Gamma_1+\Delta^2/(\omega+i\Gamma_0)$\cite{PHENOM}.  $\Gamma_1$
is a crude (i.e., frequency independent) approximation to the single particle
scattering rate, which
is extremely large in the normal state, and collapses precipitously in
the superconducting state, strongly suggesting that electron-electron
interactions determine the spectral lineshape.  $\Gamma_0$ describes the
filling in effect, and is found to be proportional to $T-T_c$, as would be
expected if it represented an inverse pair lifetime.  $\Delta$,
the gap parameter, is found to be essentially temperature
independent for underdoped samples, as was infered earlier from specific
heat data\cite{LORAM} and also observed by STM\cite{FISCHER}.
$T^*$ is then simply the temperature at which
$\Gamma_0$ becomes comparable to $\Delta$, and thus does not represent
a mean field transition temperature, despite its correlation with
$\Delta$\cite{SNSEXP}.  Similar modeling has been found to
describe STM\cite{FRANZ} and c-axis optical conductivity\cite{CARDONA} data.
This indicates that these probes are weighted towards the patch regions
of the Brillouin zone.

The above behavior can be contrasted with what occurs on the Fermi ``arcs",
which are the Fermi contour segments which connect the patches.  On the
arcs, $\Delta$ closes in a more or less BCS like fashion\cite{PHENOM},
and this occurs somewhat above $T_c$ for underdoped samples.
This behavior is found in the patch region only for overdoped samples.
It is still an unresolved
question experimentally whether $\Delta$ collapses at the same temperature
along the entire arc, though this is a distinct possibility.  It is interesting
to note that the temperature at which this collapse occurs is similar to
the temperature at which the high frequency superfluid response
vanishes for an underdoped sample with a similar $T_c$\cite{OREN}.  This is
consistent with the fact that the low energy states in the patch region are
dispersionless; that is, the patches probably do not contribute to the
superfluid density.  Therefore, once $\Delta$ collapses on the arcs, the
system no longer exhibits a superfluid response, even at high frequencies.
The available ARPES data also point to an increase of the size of the patch
region
as the doping is reduced, with a consequent decrease in the length of the
arcs.  Once the patches ``grab up" the entire zone, the metal collapses
into the Mott insulating phase.

The novelty of these findings cannot be overemphasized.  If one looks at
the anisotropy of the gap at low temperatures, it traces out a d-wave form.
This occurs even for underdoped samples\cite{MESOT}.  That is, it appears
as if there is only a single gap in the problem.  Above $T_c$, though, there
appears as if there are two gaps in underdoped samples, a more or less
conventional superconducting gap along the Fermi arcs, and a temperature
independent one in the patch regions.  Yet, at any given {\bf k} point,
the gap smoothly evolves with temperature, even through $T_c$.  This behavior
is impossible to understand from a mean field viewpoint, and simply cannot
be fixed up by attributing the gap in the patch region to some other
effect (SDW, CDW, etc.).  This is reinforced by the fact that once the gap
collapses on the arcs, the resulting Fermi surface is just a set of
disconnected segments, and therefore not derivable from a mean field
description which would have implied a continuous Fermi contour in
momentum space\cite{NAT98}.

This brings up the important question of what theoretical implications these
findings have.  A phenomenological model based on the arc-patch picture has
been developed by Geshkenbein, Ioffe, and Larkin\cite{GIL}.  The results
differ significantly in the pseudogap phase above $T_c$ from standard models
of superconducting fluctuations, since the patch regions, because of their
dispersionless nature, do not support classical fluctuations.  Their model
provides a good framework for addressing various data in the underdoped regime.
Another
phenomenological approach has been advocated by Lee and Wen\cite{LEE}.
They observe that the superfluid density can be written in the form
$\rho_s(T)=\rho_s(0)-\alpha T$, with the second term due to quasiparticle
excitations
about the d-wave nodes.  As $\alpha$ is proportional to the inverse of
$\Delta$, and $\rho_s(0)$ is proportional to $x$ (the number of doped
holes), then assuming that $\Delta$ is roughly independent of doping,
one obtains the result that $T_c \sim x\Delta$ in the underdoped
regime.  As a consequence, {\bf k} points where $\Delta_{\bf k} > \sim T_c$
have spectral gaps which survive above $T_c$.  That is, one has gapless
Fermi arcs with gapped patch regions.  It is interesting that this argument,
though completely consistent with ARPES data, is made independent of such data.

On a microscopic level, the Lee-Wen picture can be motivated by a gauge
theory approach to the t-J model\cite{WEN}.  This picture implies a d-wave
gap at low T, but a collapse of the gap along the arc above $T_c$, very
similar to what is seen by ARPES.  The disconnected Fermi arcs are a
consequence of strong fluctuations given the near degeneracy such a model
predicts between the d-wave and flux phase states.  The doping dependence
of the arcs has been calculated in a related gauge theory approach\cite{DAI}
and is
consistent with the ARPES doping trend discussed above.  A different approach
for describing the arc-patch behavior based on the t-J model has been offered 
by Furukawa, Rice, and Salmhofer\cite{FURU}.  Their results, utilizing a
renormalization group treatment of interactions in the patches, suggest a phase
separation in momentum space, with an insulating spin liquid phase in the
patch regions connected by Fermi arcs, and is motivated by studies
of three leg ladders.  Finally, there has been a recent study by Putikka,
Luchini, and Singh\cite{BILL} where the momentum distribution, $n_{\bf k}$,
for the t-J
model has been obtained by high temperature expansion.  The resulting contour
plots show that the sharp drop in $n_{\bf k}$, which defines Fermi crossings,
is only well defined on the arcs, very reminiscent of the ARPES data.

In this connection, the work of Khodel and Shaginyan deserves
mention\cite{KHODEL}.  These authors discovered another solution besides
the classic Landau Fermi liquid one.  This solution is characterized by a
flat band which is pinned to the chemical potential, either along a line, or
in an entire region of the zone.  This Fermion condensate tends to be
realized by interactions which are long range in real space and
attractive\cite{NOZIERES},
and as a consequence is unstable to superconductivity at low temperatures.
In the resulting superconducting state, dispersionless quasiparticle peaks
are found, much like what is seen by ARPES in the region of
the zone surrounding $(\pi,0)$\cite{NORM97}.
Above $T_c$, one expects very broad spectral lineshapes\cite{NOZIERES}, again
consistent with ARPES results.
In addition, $n_{\bf k}$ is predicted to have a ramp-like behavior for states
inside of the Fermion condensate (patch).  This behavior is descriptive of
that observed for the frequency integrated ARPES weight along the
$(\pi,0)-(\pi,\pi)$
direction.  Therefore, the Khodel state may indeed be the appropriate non Fermi
liquid reference state for the patch regions of the zone.

The author thanks his collaborators, Juan Carlos Campuzano, Hong Ding, and
Mohit Randeria, for the many discussions which led to the ideas presented
here.  Also, the author thanks Victor Khodel and Grisha Volovik for
discussions concerning the Khodel state.
This work was supported by the U. S. Dept. of Energy,
Basic Energy Sciences, under contract W-31-109-ENG-38.

\end{document}